\documentstyle[twocolumn,aps]{revtex}
\begin {document}
\draft
\title{Comment on ``Two-dimensional charged-biexciton complexes''}
\author{K\'alm\'an Varga}
\address{
Institute for Physical and Chemical Research (RIKEN), Wako, Saitama
351-01, Japan \\
Institute of Nuclear Research of the Hungarian Academy of Sciences 
(ATOMKI), Debrecen, H-4001, Hungary}
\date{\today}

\maketitle

\begin{abstract}
We re-examine the results of A. Thilagam 
[Phys. Rev. B {\bf 55}, 7804 (1997)],
who calculated the ratio of the binding energies of 
the two dimensional exciton complexes. We point out that, 
for charged biexciton complexes, this calculation is oversimplified, and these
systems do not always form bound states if they interact via pure Coulomb 
interaction. If experiment proved the existence of 
these objects, their bindings would 
have to be explained by other mechanisms.
\end{abstract}

\narrowtext

In a recent paper [1] A. Thilegam 
studied the binding energies of the
charged excitonic complexes in two dimensions. The author derives an 
analytical
expression for the binding energy of the charged exciton, which is an
AAB-type Coulomb three-body system, and estimates the binding energy 
ratio of the charged exciton and 
the exciton (AB-type Coulomb two-body system) as a function of the mass ratio
of the heavier to the lighter constituent. 
The analytical formula is based on the assumption that the optimum 
binding in two dimensions is provided by a configuration in which
the particles are placed along a line (eq. (13) of Ref. [1]).
The result of this formula agrees  to some extent 
with the variational calculation  of Stebe and Ainane [2]. 
\par\indent
To extend the approach  to AAABB-type five-body Coulomb systems, 
it is argued that this five-particle
system can be effectively substituted by an AAB-type three-body system.
The reason, according to the author, is that an  AABB system ( 
say,  an  H$_2$ molecule) reduces to an  AB system 
(a H atom) in two dimensions. 
To justify this crucial assumption the author refers to a recent paper
 [3]. The authors of Ref. [3] state  that 
in two dimensions  the particles in an  AABB-type 
four-body  system form a square, and, as a consequence, 
the biexciton Hamiltonian is reduced to that of the exciton. The questions 
are, however: Does the AABB-type biexciton form a square-like equilibrium
configuration inside an AAABB-type charged biexciton as well? And, if so, 
where is the fifth particle? In the five-particle  case, 
can one also reduce the Hamiltonian of four of the five  
particles into that of an exciton? On what grounds  
can one choose two A-type particles from the three to participate in 
the reduction of the AABB Hamiltonian? 
\par\indent
Based on the above assumption, the author applies the AAB  
binding energy formula to AAABB-type charged biexciton systems, and plots
the ratio of  the binding energies  of the AAABB and  AAB 
charged exciton complexes for both positively and negatively charged systems.
\par\indent
In this Comment  we would like to point out that these  
assumptions are oversimplified, and the results for charged biexcitons
are not correct. We question the validity of this approximation because, 
as will be shown in the 
following, the neglected antisymmetrization, exchange interactions
and dynamical effects play vital roles in these systems. 
\par\indent
Unlike in the case of the AAB-type three-body system, the wave function of the
AAABB-type five-body system has to be (anti)symmetrized explicitly. 
Even if A is a fermion, explicit antisymmetrization in the AAB-type system 
may be avoided. Then the model describes a situation in which the subsystem 
AA forms a spin singlet, and the spatial part of the wave function is 
just symmetrical. But in the five-body case we have three 
indistinguishable particles A. By neglecting the 
(anti)symmetrization, the formula is assumed to be equally applicable
to bosons or fermions. That this is not correct is shown by the fact that, 
for the mass ratio $\sigma\equiv m_A/m_B=1$,  
the bosonic and fermionic AAABB-type systems are found to be rather 
different in 3D [4]. 
The bosonic five-body Coulomb system is bound with a 
relatively large binding energy, but no bound state exists  for 
fermions. 
\par\indent
To show that the neglected dynamics plays an essential role, we present 
a variational solution to the five-body problem and compare 
with that obtained by the binding energy formula of Thilagam 
[1]. 
The variational solution 
is calculated by the stochastic variational method [4]. This method 
has been tested against several benchmark tests and it has been demonstrated 
to give very accurate results for N-particle systems (N=3,4,5,6). 
The Coulombic systems for which we tested our method against others are 
H$_2$, H$_2^+$, H$^-$ in 3D and the positronium molecule in 2D and 3D. 
Our calculations proved to be very accurate in all cases [4].
\par\indent
Table I shows the results of our variational calculations for the 
two-dimensional exciton complexes for $\sigma=0$ and for $\sigma=1$. 
For the three-body cases our results are in good agreement with those 
of Refs. [1,2]. For example, in the case of H$^-$ (AAB-type, $\sigma=0$)
the binding energy ratio is predicted to be 0.5 by ref. [1], 0.48 
by Ref. [2] and 0.48 by the present calculations. For the $\sigma=1$ case the 
corresponding ratios are 0.2875 (Ref. [1]), 0.24 (Ref. [2]) and
0.24 by the present calculation. In the case of H$_2^+$ (ABB-type system 
with $\sigma=0$), however, our 
result (1.64) disagrees with that of Ref. [1] (1.25) (this case is not 
included in Ref. [2]).
\par\indent
In our calculation, the binding energy ratio of the positively 
charged biexciton with respect to the 
positively charged exciton, for $\sigma=0$, is found 
to be 1.32. 
This value is much larger than  
the prediction of Ref. [1] (0.17). Note that the 
corresponding ratio (the binding energy of H$_3^+$ divided
by the binding energy of H$_2^+$) is about 1.7 in 3D. This binding strongly 
decreases with increasing $\sigma$ (this is just the opposite
of the behaviour of the curve in  Fig. 4 of Ref. [1]), 
and the positively charged biexciton becomes unbound around $\sigma=0.2$. 
We have not found bound systems of negatively 
charged biexcitons with total orbital momentum $L=0$, therefore, we 
cannot support the existence of the upper curve of Fig. 4 of Ref. [1].
\par\indent
As a consequence, we think that the conclusion of Ref. [1] that 
``the charged biexciton complex is stable for any mass ratio $\sigma$'' is 
unfounded. It is, of course, difficult to prove the nonexistence 
of something; we only state that Ref. [1] did not prove the existence. 
The bound-state nature of a state can only be 
proved by calculating the total energy and comparing it to the relevant
threshold, which, in this case, is the energy of the 
biexciton. The binding 
energy formula of Ref. [1] provides binding energy ratios without reference to
the energy of the biexciton, therefore, it provides no information 
on the stability. 
\par\indent
In conclusion, to determine the binding energy of 
biexciton complexes and to find their stability regions, more elaborate
calculations are needed. If measurements were to show the 
existence of charged biexcitons, then one would have to check the effects
of other mechanisms that may increase the binding energies, for example, 
magnetic field or localization (confinement) effects.

\bigskip
\par\indent
This work was supported by OTKA grant No. T17298 (Hungary). The author
is gratefully thanks Prof. R. G. Lovas for careful reading of the
manuscript.

\begin{table}

\caption{Energies ($E(\sigma)$) and binding energies 
with respect to the relevant thresholds  ($B(\sigma)$)
of the 2D exciton complexes (in a.u.). Particle A
carries negative and particle B carries  positive 
charge. ($\sigma=m_A/m_B$). The asterisk expresses 
that the state is found to be unbound.}
\begin{tabular}{ccccc}\hline
system & $E(\sigma=0)$  & $E(\sigma=1)$ &  $B(\sigma=0)$  & $B(\sigma=1)$ \\
\hline
AB    & $-2.00$      & $-1.00$   & $-2.00$      & $-1.00$     \\
AAB    & $-2.24$      & $-1.12$  & $-0.24$      & $-0.12$     \\
ABB    & $-2.82$      & $-1.12$  & $-0.82$      & $-0.12$     \\
AABB   & $-5.33$      & $-2.19$  & $-1.33$      & $-0.19$     \\
AAABB  & $  *  $      & $ *   $  & $ 0.00$      & $ 0.00$     \\
AABBB  & $-6.41$      & $ *   $  & $-1.08$      & $ 0.00$     \\
\end{tabular}

\end{table}


\begin{thebibliography}{99}
\bibitem{Thilagam} A. Thilagam 
Phys. Rev. B {\bf 55} 7804, (1997).
\bibitem{variational} B. Stebe an A. Ainane, Superlatt. Microstuct.
{\bf 5}, 545 (1989).
\bibitem{singh} J. Singh, D. Birkedal, V. G. Lyssenko, and J. M. Hvam,
Phys. Rev. B {\bf 53}, 15909 (1996).
\bibitem{SVM} K. Varga and Y. Suzuki, Phys. Rev. C {\bf 52}, 2885 (1995);
Phys. Rev. A {\bf 53}, 1907 (1996).
\end{thebibliography}
\end{document}